\documentclass[aip,jap,reprint,groupedaddress,floatfix]{revtex4-1}

\usepackage{graphicx}
\usepackage{hyperref}
\usepackage{amsmath}
\usepackage{wasysym}
\usepackage{amssymb}

\begin{document}

\author{Tanuj Trivedi}
\email{tanuj@utexas.edu}
\author{Sushant Sonde}
\author{Hema C. P. Movva}
\author{Sanjay K. Banerjee}
\email{banerjee@ece.utexas.edu}
\affiliation{Microelectronics Research Center, The University of Texas at Austin, Austin, TX 78758, United States}

\title{Weak Antilocalization and Universal Conductance Fluctuations in Bismuth Telluro-Sulfide Topological Insulators}


\begin{abstract}
We report on van der Waals epitaxial growth, materials characterization and magnetotransport experiments in crystalline nanosheets of Bismuth Telluro-Sulfide (BTS). Highly layered, good-quality crystalline nanosheets of BTS are obtained on SiO$_2$ and muscovite mica. Weak-antilocalization (WAL), electron-electron interaction-driven insulating ground state and universal conductance fluctuations are observed in magnetotransport experiments on BTS devices. Temperature, thickness and magnetic field dependence of the transport data indicate the presence of two-dimensional surface states along with bulk conduction, in agreement with theoretical models. An extended-WAL model is proposed and utilized in conjunction with a two-channel conduction model to analyze the data, revealing a surface component and evidence of multiple conducting channels. A facile growth method and detailed magnetotransport results indicating BTS as an alternative topological insulator material system are presented.
\end{abstract}


\maketitle
 \vspace{-1em}
\section{Introduction}
\label{sec-intro}
Bismuth (Bi) and Antimony (Sb) based binary chalcogenide compounds have received widespread attention in the last few years for exhibiting rich physics of strong 3D topological insulators (TI), matching theoretical predictions of time-reversal invariant and gapless surface states with spin-momentum locked Dirac fermions \cite{FuKane07,Hasan_Kane_10,Ando13}. Applications of TI heterostructure devices range from topological quantum computing with Majorana fermions\cite{Hasan_Kane_10}, spin-based logic and memory\cite{Pesin_Spin_12} and axion electrodynamics\cite{Li_Axion_10}. Electronic and spintronic device applications of TI often involve manipulating the electronic surface states, and hence uncovering the details of the underlying transport mechanism is an important aspect of current research. The binary phases (Bi, Sb)$_2$(Se, Te)$_3$ have been explored extensively as 3D TI materials from the standpoint of transport experiments\cite{Chen_WAL_10,Steinberg_SS_10,He_BSMBE_11,Wang_EEI_11,Liu_EEI_11,Chen_aWAL_11,Steinberg_WAL_11,Bansal_TI_12,Taskin_BS_12,Takagaki_ST_12,Roy_BT_13,Ando13}. Transport experiments often involve the non-negligible contribution from the bulk, which complicates the electronic probing of surface states. Attributing the indirect signatures to multiple conduction channels has been the focus of intense research\cite{Ando13,Chen_aWAL_11,Steinberg_WAL_11,Taskin_BS_12,Bansal_TI_12,Lee_BSTS_12,Kim_Coherent_13,Yang_BSTGating_14}, prompting the need to further explore transport in multiple TI material systems. An attractive direction to explore is ternary (or quaternary) compounds M$_2$X$_{3-x}$Y$_x$ (M = \{Bi, Sb, Bi$_{1-y}$Sb$_y$\}, X, Y = \{Te, Se, S\}), as the material properties, and consequently transport properties, can be tuned by the addition of other elements\cite{Ren_BST_10,Wang_Ternary_11,Kong_BST_11,Lee_BSTS_12,Xia_BSTS_13,Ando13}. The Sulfur-based ternary compound (naturally occurring tetradymite with an ideal formula Bi$_2$Te$_2$S) has received relatively little attention from a transport perspective, even as it is theoretically predicted to be a promising 3D TI\cite{Wang_Ternary_11}. The Sulfur based tetradymite has been synthesized in the laboratory as bulk crystals in previous experiments, albeit showing non-stoichiometry in deviation from the ideal structure\cite{Soonpaa_BTS_62,Glatz67,Pauling75,Ji_BTS_12}. The substitution of a more electronegative S for Te in the Bi$_2$Te$_3$ crystal structure is expected to increase the bulk band gap, reduce antisite defects and exposes the otherwise buried Dirac point\cite{Wang_Ternary_11,Ji_BTS_12}, as has been observed in promising angle-resolved photoemission spectroscopy (ARPES) experiment on bulk crystals of the material\cite{Ji_BTS_12}. The tetradymite ternary thus provides a promising platform to study transport signatures of the surface states.\par We report on the van der Waals epitaxial growth of crystalline Bismuth Telluro-Sulfide (BTS) nanosheets and observation of surface states through transport experiments. Low-temperature insulating ground state in the conductivity of the BTS devices reveals the presence of electron-electron interaction (EEI) effects, which have been observed for thin films of 3D TIs. The characteristic weak antilocalization (WAL) signature of the topological surface states is seen in the magnetoresistance, which acts in combination with EEI effects at low-temperatures and low-fields. Evidence of separation of transport channels in Hall data is seen, with a parallel conductivity contribution from bulk states. A two-channel Hall conductance model is used in conjunction with an extended WAL fit to describe the results. Universal conductance fluctuations (UCF) are also observed in the magnetoresistance of thin BTS devices, the temperature-dependent behavior of which is analyzed with standard UCF theory for two-dimensional metals, yielding phase coherence lengths of the same order as those obtained from WAL. Empirical parameters from modeling the thickness- and temperature-dependent WAL and UCF data indicate two-dimensional mesoscopic transport, revealing the presence of accessible surface states in the BTS material system.
\vspace{-1em}
\section{Experimental Methods}
\label{sec-methods}
\subsection{van der Waals epitaxial Growth}

The BTS nanosheets are grown by hot-wall van der Waals epitaxy (vdWE) with a combination of compound solid-state precursors, in a 60mm diameter quartz tube inside a three-zone furnace. Low-resistivity ($\sim$ 5 $m\Omega$cm) silicon wafers are thermally oxidized to grow high quality $285-300$ nm thick SiO$_2$. An alignment marker grid for e-beam lithography is then etched into the SiO$_2$ film with standard photolithography and dry etch, instead of depositing metallic markers. Metal alignment markers are found to act as nucleation centers leading to undesirably thick, dense and malformed growth with possible metal contamination, hence etched-in alignment markers are preferred. SiO$_2$/Si wafers so prepared are then cleaved into samples of 5-20 mm size and placed either vertically or horizontally in a slotted quartz carrier. The quartz carrier is placed at the neck of the furnace in the cold-zone, downstream of the precursor materials ($\sim$16" away from central zone). Muscovite mica samples of similar sizes are freshly cleaved immediately prior to growth and loaded inside in a similar fashion. Powdered Bi$_2$Te$_3$ (4N Sigma-Aldrich) is placed in a quartz boat in the center zone, along with chunks of Bi$_2$S$_3$ (5N Sigma-Aldrich), either in the same boat or in another boat in the zone closer to the sample carrier. The quartz tube is then pumped down to base pressure several times and subsequently purged with N$_2$ gas for a few hours to remove any trace oxygen and moisture, and to achieve a stable pressure and flow of the carrier gas. All three zones of the furnace are then heated up to $510^\circ$C within 20 minutes without overshooting and are held at that temperature for $20-40$ minutes before being cooled down naturally to room temperature. Good growth is observed when tube pressure is in the range of $20-100$ Torr with N$_2$ flow in the range of $150-200$ sccm and when the samples are in the temperature range of $360-380^\circ$C. Representative samples are analyzed with Raman spectroscopy and X-ray diffraction to confirm crystalline nature of the growth. Compositional analysis is performed on candidate nanosheets with Carl Zeiss/EDAX energy dispersive x-ray spectroscope to confirm the presence of all three elements within a range of stoichiometries. Tellurium-rich nanostructures are obtained for sample temperatures lower than $\sim 250^\circ$C, as has been observed before in a similar growth experiment\cite{Takagaki_HWE_11}.
\vspace{-1em}
\subsection{Device Fabrication}
As-grown samples of BTS on SiO$_2$ are inspected using an optical microscope and AFM to identify flat and thin ($\sim 7-100$ nm) candidate nanosheets, with lateral dimensions in the range of a few microns. Contacts are patterned directly on the as-grown nanosheets in a four-point or Hallbar geometry with standard e-beam lithography and liftoff. Immediately prior to metallization, the contact areas on the BTS nanosheet are etched with a brief Ar plasma ($\sim$10-12 s, 75 W) in an RIE chamber using the e-beam resist as the etch mask, to remove surface oxides and other contaminants. A $3/30$ or $5/120$ nm of Ti/Pd or Ti/Au metal stack is deposited with e-beam evaporation for the contact leads. The samples are then attached onto a standard 16-pin DIP package with silver-paste and wirebonded using a Au ball-bonder or an Al wedge-bonder. Sample temperature during the fabrication process is carefully maintained below $150^\circ$C to prevent material degradation.
\vspace{-1em}
\subsection{Transport Measurements}
The wirebonded samples are loaded inside Quantum Design EverCool2 PPMS system, equipped with a $9$ T magnet. All magnetotransport measurements are performed using Stanford Research Systems 830 digital lock-in amplifiers. A steady current in the range of $0.1-1 \mu A$ is supplied to the two outer terminals of the Hallbar. The current source is formed by the sinusoidal voltage output of the SRS-830 and a standard series resistor. The series resistor is in the range of $\sim$1 M$\Omega$, whereas the typical DUT resistances are of the order of $\sim$1 k$\Omega$ or less. Hence, the current fluctuation due to the DUT series-loading is three orders of magnitude lower and can be safely ignored. Four-point longitudinal ($r_{xx}$) and transverse ($r_{xy}$) resistances are measured as a function of the magnetic field $B$, with two phase-locked lock-in amplifiers at low frequency ($11-13$ Hz). The symmetric $R_{XX,\ XY}(B)$ functions are calculated as: $R_{XX}(B) = \frac{1}{2}\cdot[r_{xx}(B) + r_{xx}(-B)]$ and $R_{XY}(B) = \frac{1}{2}\cdot[r_{xy}(B) - r_{xy}(-B)]$. Temperature dependent measurements are performed down to a chamber temperature of 2 K, and magnetic field sweeps are up to $\pm 9$ T.
\vspace{-1em}
\section{Results and Discussion}
\label{sec-results}
\subsection{Growth, Structural and Chemical Characterization}
\label{sec-growth}
			\begin{figure}[h]
			\centering
				\includegraphics[width=0.5\textwidth]{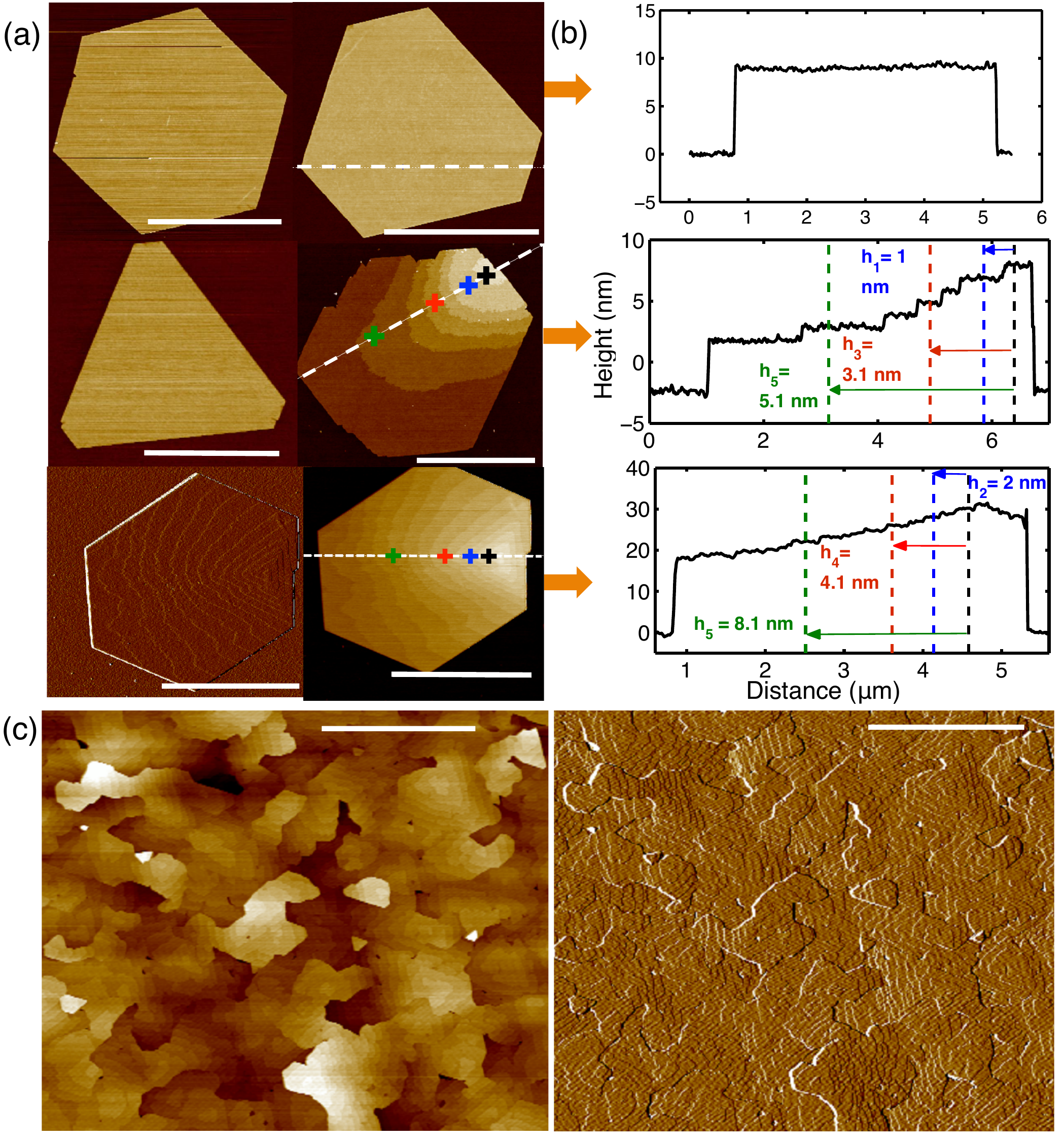}
				\caption{(a) AFM images of representative as-grown BTS nanosheets (scale bars $3\mu$m) on SiO$_2$. (b) Cross-sectional height profiles at the dashed lines showing examples of a flat surface (top) and layered growth terraces (mid and bottom). (c) AFM height (left) and amplitude error (right) plots for a continuous dense growth on mica substrate, showing the layered or terraced growth of the nanosheets.}
			\label{fig-AFM}
			\end{figure}

The ability to grow any layered material on top of any other layered or a 3D bulk material\cite{Koma99} makes vdWE a versatile and convenient method to synthesize novel materials. Bi and Sb chalcogenide compounds have been grown using vdWE on 3D substrates\cite{Ferhat96,Kong_BST_11,Takagaki_HWE_11,Hao_vdWE_12}, layered substrates such as mica\cite{Peng12}, hexagonal Boron Nitride\cite{Gehring_vdWE_12} and graphene\cite{Dang10}. The hot-wall vdWE technique is preferable for growing thin-films of TI compounds containing Sulfur, as the high vapor-pressure of Sulfur makes it an undesirable source material for most molecular beam epitaxy (MBE) systems. The vdWE-grown BTS nanosheets show clear crystal shape-symmetry, growing largely in hexagonal and truncated-triangular shapes of lateral dimensions of a few microns, as seen in Fig-\ref{fig-AFM}(a). The underlying crystal symmetry of BTS is trigonal-hexagonal (space group $R\bar{3}m$), which leads to the formation of layered triangular nanosheets during growth. Similar terraced growth has also been observed for other 2D material systems with trigonal symmetry, on different substrates\cite{Michely_Pt_93,Roy_BT_13,Roy_CrTe_15}. After the initial nucleation of the islands, the ultimate shape is dependent on the variance in the growth rate along the different types of edges in the hexagonal-trigonal shape, which has been established by Monte Carlo simulation of the kinetic growth mechanism\cite{Michely_Pt_93,Liu_IslandGrowth_93}, leading to either hexagonal or more often truncated-triangular nanosheets. The vdWE-grown BTS nanosheets are found to nucleate randomly on the SiO$_2$ surface, but show evidence of highly layered growth, visible in the atomic force microscopy (AFM) height profiles. Two such examples are shown in Fig-\ref{fig-AFM}(a): a height plot (mid-right) and an amplitude error (bottom-left) and height plots (bottom-right). Fig-\ref{fig-AFM}(b) mid and bottom figures show the step height profiles, measured between subsequent terraces, accurately match the quintuple unit cell thickness of $\approx 1$ nm. The BTS unit cell is also a quintuple layer similar to that of Bi$_2$Te$_3$, characteristic of the tetradymite crystal structure. BTS nanosheets also grow in a similar fashion on mica substrates, sometimes showing densely nucleated terraced growth as seen in Fig-\ref{fig-AFM}(c). Denser growth on mica is favorable, as the freshly cleaved mica substrate is very highly crystalline in-plane with trigonal-hexagonal symmetry, which acts as a template for the denser nucleation. 

			\begin{figure}[h]
			\centering
				\includegraphics[width=0.5\textwidth]{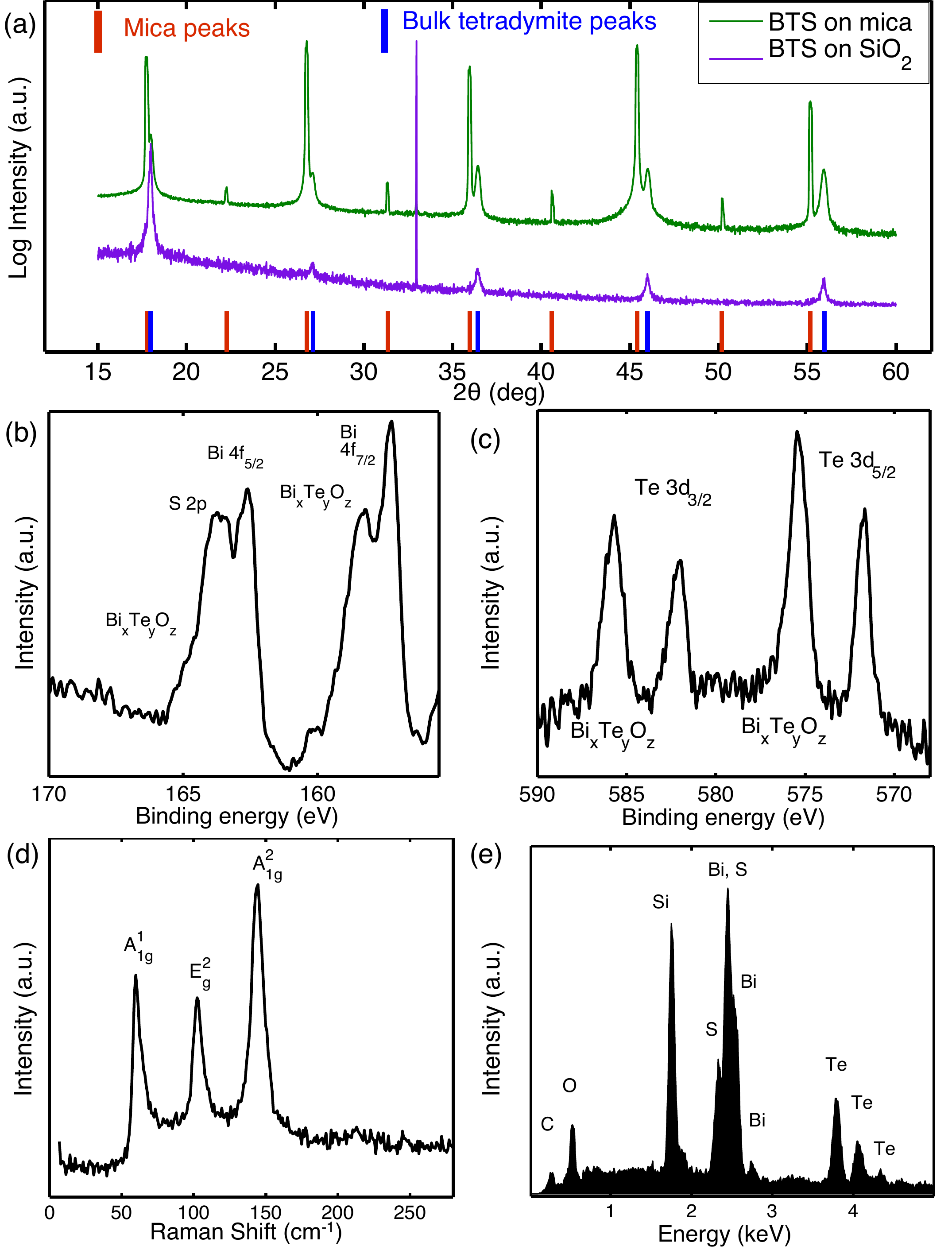}
				\caption{(a) X-ray diffraction pattern from as-grown BTS on mica and SiO$_2$ substrates, compared with bulk crystal (0 0 n) peak locations. (b), (c) Core-level X-ray photoelectron spectra showing the Bi 4f, S 2p (b) and Te 3d (c) bonding states in a candidate BTS sample, also showing the presence of a surface oxide. (d) Raman shift spectrum measured from as-grown candidate BTS nanosheet. (e) Energy dispersive X-ray analysis on as-grown BTS nanosheet showing the presence of Bi, S and Te elements.}
			\label{fig-MatChar}
			\end{figure}
			
X-ray diffraction (XRD) spectra of as-grown BTS-on-SiO$_2$ and BTS-on-mica samples show evidence of particularly c-axis oriented growth, i.e., sharp peaks only at the positions of the (0 0 n) facet reflections of the bulk tetradymite crystal\cite{Glatz67,AM_RRUFF,Ji_BTS_12}, as seen from Fig-\ref{fig-MatChar}(a). The AFM and XRD results confirm that, once nucleated, the BTS nanosheets subsequently grow epitaxially, forming the layered van der Waals crystal structure of the tetradymite. Raman spectra show sharp peaks (see Fig-\ref{fig-MatChar}(d)), which are coincident with tetradymite spectrum\cite{AM_RRUFF,Chis_raman_12}. Major Raman shifts are observed at A$_{1g}^1\sim61-63$ cm$^{-1}$, E$_g^2\sim103$ cm$^{-1}$ and A$_{1g}^2\sim144$ cm$^{-1}$, which when compared to Bi$_2$Te$_3$, exhibit a blue shift. The S-atoms occupy the middle chalcogen layer and intermix with Te-atoms in the outer chalcogen layers in the BTS crystal\cite{Pauling75,Ji_BTS_12}. Replacing the Te atoms in the Bi$_2$Te$_3$ unit cell with the smaller S atoms leads to a smaller Bi$-$S bond length, more compressive strain and non-stoichiometry in the ideal tetradymite lattice, leading to the so called $\gamma$-phase in the temperature range of the growth ($360-380^\circ$)\cite{Glatz67,Pauling75,Ji_BTS_12}. The compressive strain leads to the blue shift observed in the Raman spectrum\cite{Chis_raman_12}. Core-level X-ray photoelectron spectroscopy (XPS) analysis on candidate BTS samples show the presence of the expected Bi 4f, S 2p and Te 3d bonding states, shown in Fig-\ref{fig-MatChar}(b) and (c)\cite{Roy_BT_13,Purkayastha_BTS_08,Sima_BSTS_12}. The chemical shifts at higher bonding energies away from the Bi 4f $5/2,\ 7/2$ states point to the presence of a surface oxide, as do the ones observed for the Te 3d $3/2,\ 5/2$ states\cite{Purkayastha_BTS_08}, justifying the need for plasma treating the surface of the BTS nanosheets before contact metal deposition during device fabrication. Energy dispersive X-ray spectroscopy on candidate BTS nanosheets on SiO$_2$ and mica show the presence of Bi, Te and S in all samples (see Fig-\ref{fig-MatChar}(e)) and stoichiometries are established in the range of Bi$_2$Te$_{2-x}$S$_{1+x}$ with $x\in[0.2,0.5]$, in near agreement with the $\gamma$-phase, the proposed alternative to the ideal structure of the tetradymite\cite{Glatz67,Pauling75,Soonpaa_BTS_62}.
\vspace{-1em}
\subsection{Temperature-dependent Conductivity Measurements}
\label{sec-RvT}
			\begin{figure}[h]
				\includegraphics[width=0.5\textwidth]{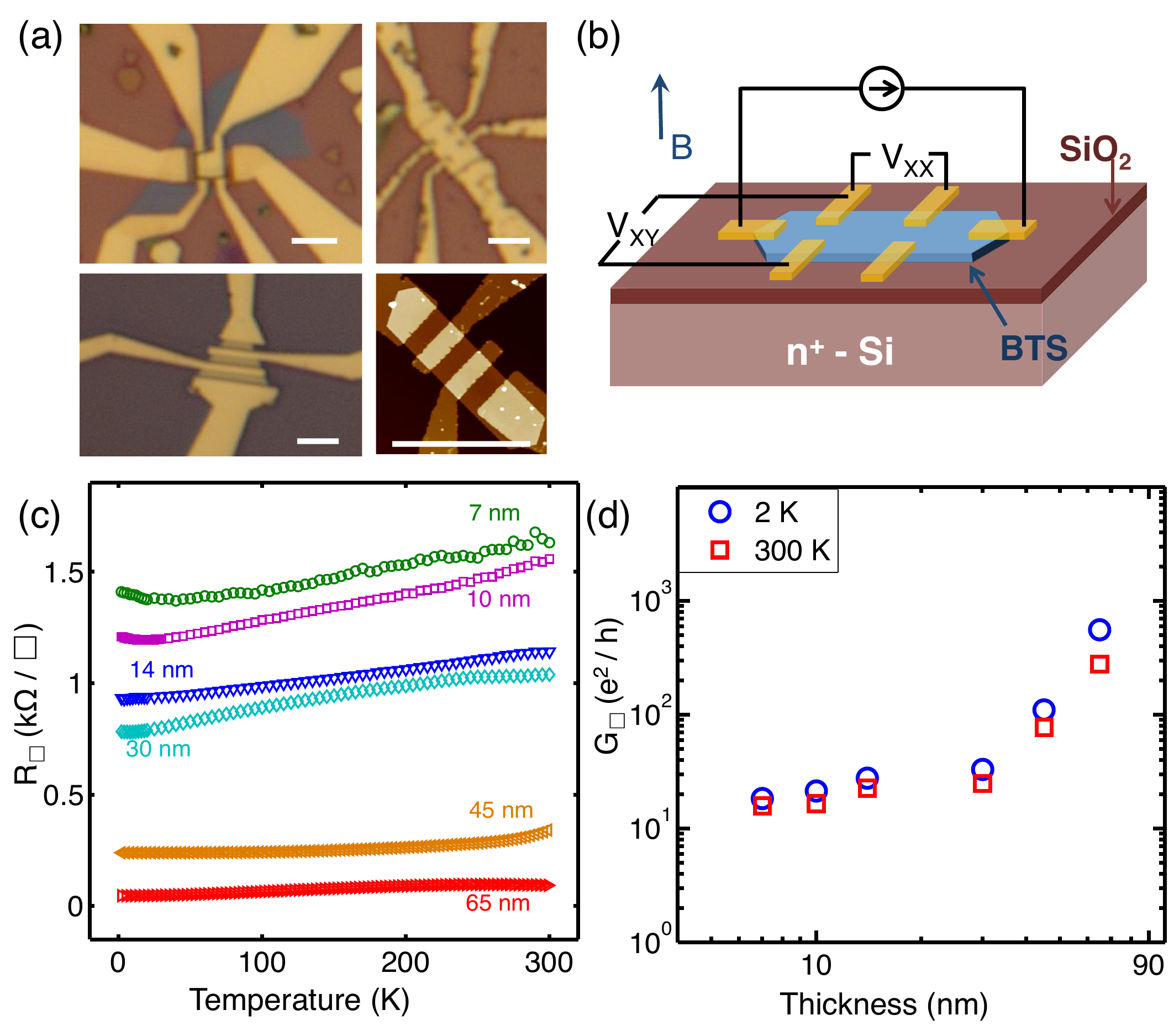}
				\caption{(a) Optical and AFM images of several devices (all scalebars are 3 $\mu$m). (b) Schematic diagram of a typical BTS device structure with the constant current source geometry. (c) Device sheet resistance measured as a function of temperature for different BTS thicknesses. (d) Device sheet conductance as function of BTS thickness $d$, in units of $e^2/h$.}
			\label{fig-RvT}
			\end{figure}
Some examples of devices fabricated on as-grown BTS nanosheets on SiO$_2$ substrates in a four-point or Hallbar geometry, are shown in Fig-\ref{fig-RvT}(a). All measurements were done in the constant current mode, as shown in the representative schematic in Fig-\ref{fig-RvT}(b). Four-point resistance was measured as a function of the sample temperature, showing an almost linearly decreasing resistance for all devices (see Fig-\ref{fig-RvT}(c)). This is indicative of a doped bulk, likely due to chalcogen deficiencies (donors) and antisite defects (acceptors), characteristic of Bi-chalcogenide TI materials\cite{Wang_defects_13}. The BTS devices were found to be dominantly n-type from Hall data. Due to a vapor pressure higher than Tellurium, Sulfur evaporates more during growth and device processing leaving behind donor vacancies, while the Bi-S bonding in BTS reduces acceptor-like antisite defect formation, leading to an overall n-type behavior\cite{Ji_BTS_12}. This observation was also made for the bulk crystal case in previous experiments\cite{Soonpaa_BTS_62,Ji_BTS_12}. Care must thus be taken to reduce the overall fabrication and processing temperature, as was done in this study. Interestingly, a previous experiment on non-stoichiometric BTS bulk crystals also showed highly anisotropic conductance, i.e., the ratio of in-plane ($\sigma_{\parallel}$, perpendicular to the c-axis of the crystal) to out-of-plane ($\sigma_\bot$, parallel to the c-axis of the crystal) conductivity was large\cite{Soonpaa_BTS_62}. Sheet conductance data for several BTS devices are also plotted vs the BTS nanosheet thickness $d$ in Fig-\ref{fig-RvT}(d). $G_\square$ is approximately flat for thinner devices, which indicates the large contribution of the surface channel to the sample conductance\cite{He_BSMBE_11,Bansal_TI_12}. As $d$ increases, a corresponding increase in $G_\square$ is seen, not unlike a doped semiconductor. This increase in conductivity indicates the growing contribution of the bulk channel for thicker nanosheets, which has been explained as increased impurity band states\cite{He_BSMBE_11}. 
			\begin{figure}[h]
				\includegraphics[width=0.5\textwidth]{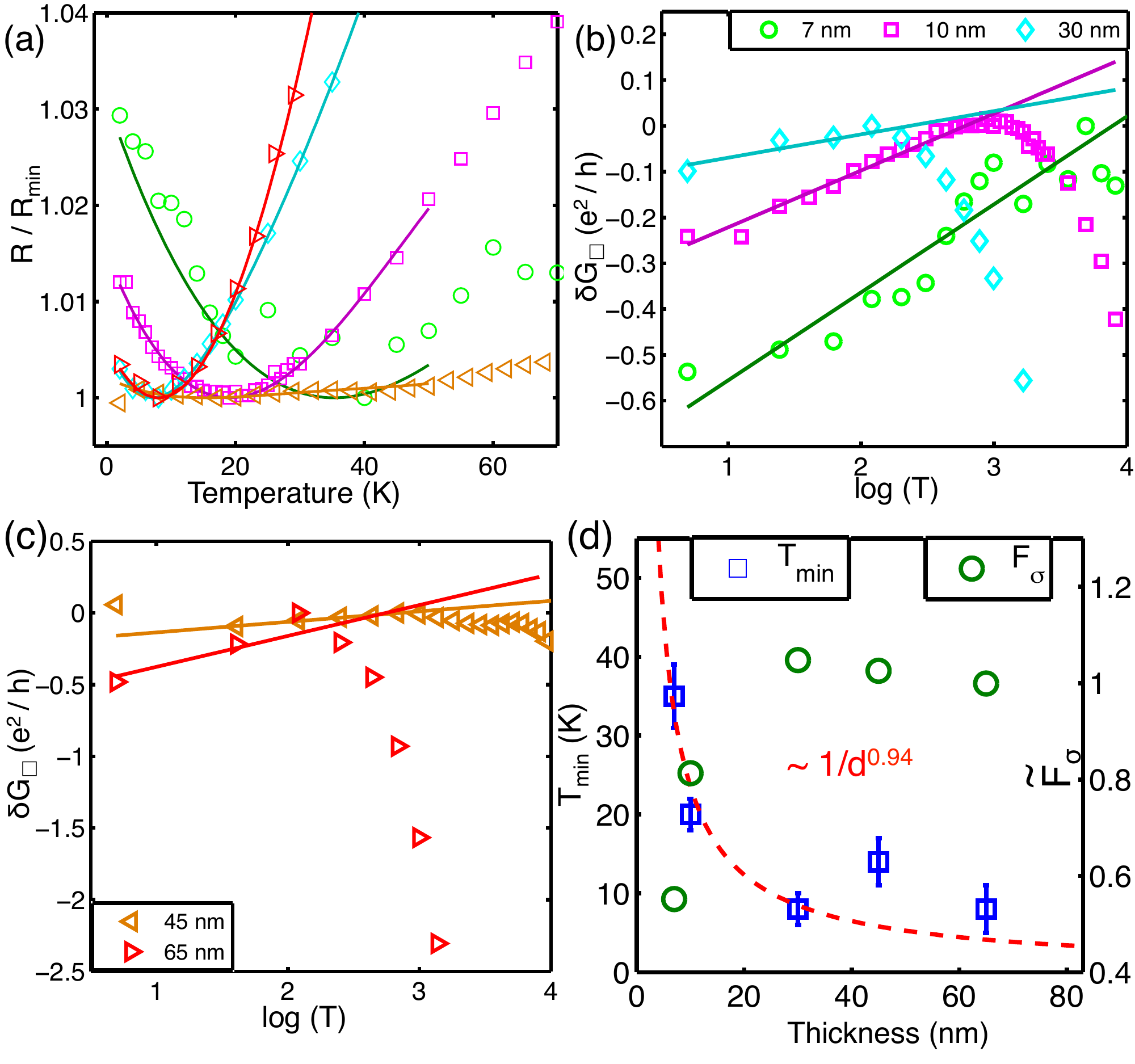}
				\caption{(a) Several normalized $R_{\square}$ vs $T$ data shown at low temperatures, to illustrate the insulating ground state. Solid lines are a guide to the eye. (b), (c) Linear fits to conductivity variation with logarithmic temperature, due to EEI effects. (d) Temperature of resistance minima ($T_{min}$) and Coulomb screening factor $\tilde{F_\sigma}$ as a function of thickness $d$. The dashed red line is a $\sim d^{ -0.94}$ fit to the $T_{min}$ data.}
				\label{fig-RvTEEI}
			\end{figure}
	\par There is a decrease in the rate of reduction of resistance at lower temperatures ($<50$ K), oftentimes showing an increase in the resistance (see Fig-\ref{fig-RvTEEI}(a)), or a decrease in conductivity. This decrease in the conductivity is linearly proportional to logarithmic temperature and is indicative of an insulating ground state, expected for a 2D system with electron-electron interaction (EEI), in which the Coulomb interaction between electrons is enhanced and becomes long range\cite{LeeRama85,Altshuler_EEIBOOK_85,Liu_EEI_11,Wang_EEI_11,Lu_EEIWAL_14}. The correction to the conductivity due to the dynamically screened interaction can be expressed as\cite{LeeRama85,Beutler_EEI_88}:
	\begin{equation}\label{eq-EEI}
		\delta\sigma_{2D} = \frac{e^2}{2\pi h}\left[2 - \frac{3}{2}\tilde{F_\sigma}\right] ln\left(\frac{T}{T_0}\right)
	\end{equation}
Where the fitting parameter $\tilde{F_\sigma}$ is a Hartree term related to the strength of Coulomb screening and $T_0$ is a reference characteristic temperature, taken as 2 K for this experiment\cite{LeeRama85,Beutler_EEI_88}. Example fits for the different BTS devices are shown in Fig-\ref{fig-RvTEEI}(b) and (c). Incidentally, the conductivity for the $d=14$ nm device was observed to flatten out at low temperatures, unlike other devices that show a decrease, and hence could not be fitted to the EEI model. The likely reason for this observation is that out of the competing contributions from WAL and EEI, the WAL contribution is larger than EEI for the 14 nm device, and hence the decrease in conductivity due to EEI is smaller than that observed for other devices. The competing contributions are discussed in more detail in Section \ref{sec-MRvT}. The exact definition of the fit parameter $\tilde{F_\sigma}$ depends on the dimensionality of the sample, which for two-dimensional films is: $\tilde{F_\sigma}^{2D} = \frac{8}{F}\cdot(1+\frac{F}{2})\cdot ln(1+\frac{F}{2}) - 4$, where $F$ is the dimensionless interaction averaged on the Fermi surface, and for values of $0<F<1,\ \tilde{F_\sigma}\sim F$ within $10\%$\cite{LeeRama85}. The nature and values of the screening parameter $\tilde{F_\sigma}$ are a matter of some debate due to the immense difficulty involved in its exact calculation and variance in experimental observation\cite{LeeRama85,Altshuler_EEIBOOK_85,Lin_SOEEI_01,Chen_aWAL_11,Takagaki_CuBS_12}. However, qualitatively from the Thomas-Fermi theory the values of $F$ from $0$ to $1$ signify weaker screening (larger correction) to stronger screening (smaller correction) for most metals in presence of disorder\cite{Lin_SOEEI_01,Lu_EEIWAL_14}. The values of $\tilde{F_\sigma}$ from the EEI fits are shown in \ref{fig-RvTEEI}(d) as a function of nanosheet thickness, going from a value of $\sim 0.56$ for thinner to $\sim 1$ for thicker devices. These values indicate stronger screening in thicker devices, likely due to larger contribution from bulk carriers to the transport, thus leading to a smaller correction to $\sigma(T)$\cite{Lu_EEIWAL_14}. Similar values in this range have been reported for devices of TI materials and thin films of strong spin-orbit coupling materials, such as elemental Bi\cite{Altshuler_EEIBOOK_85,McLachlan_Bi_83,Woerlee_Bi_83,Liu_EEI_11,Wang_EEI_11,Chen_aWAL_11,Lu_EEIWAL_14}.
The temperatures $T_{min}$, when $R = R_{min}$, are plotted vs thickness ($d$) of the BTS nanosheet in Fig-\ref{fig-RvTEEI}(d) showing a $\sim d^{-0.94}$ fit. This $\sim 1/d$ behavior of $T_{min}$ can be derived from a conduction model considering both surface and bulk channels contributing to the total transport (see Appendix \ref{appsec-EEI} for derivation). The decreasing $T_{min}$ vs $d$ data are qualitatively similar to the observation made for the sheet conductance vs $d$, as the bulk channels become more dominant for thicker devices and the onset of the 2D interaction-driven ground state is evident at lower temperatures.
\vspace{-1em}
\subsection{Thickness-dependent Magnetotransport Measurements}
\label{sec-MRvd}
			\begin{figure}
				\includegraphics[width=0.5\textwidth]{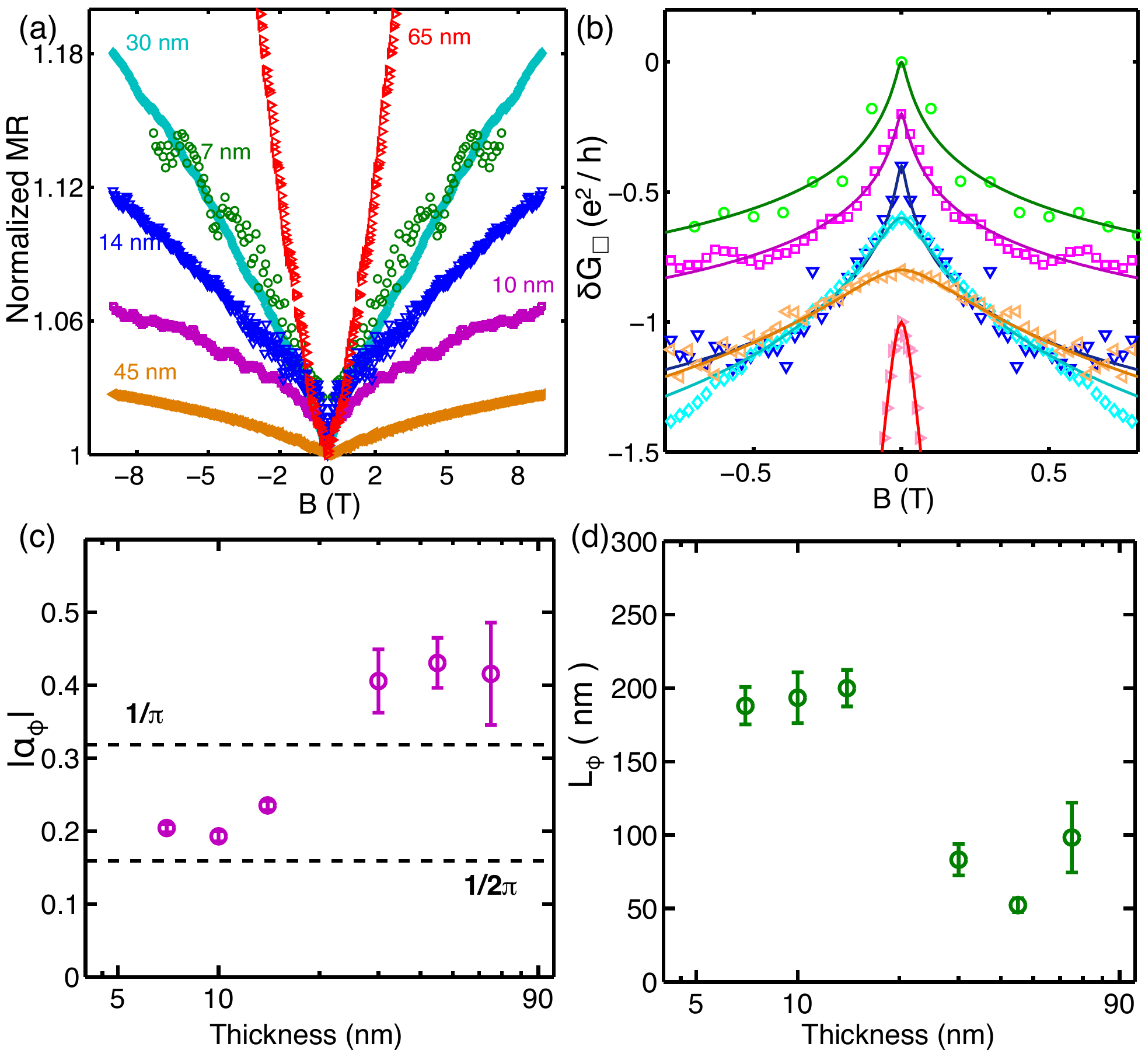}
				\caption{(a) Normalized symmetric longitudinal magnetoresistance $R_{XX}$ for several devices at 2 K. (b) Reduced HLN fit to $\delta G_{\square}$ at the low-field WAL feature (solid lines are fits). Curves are shifted for clarity. (c) Reduced HLN prefactor $\alpha_\phi$ vs BTS thickness $d$. (d) Reduced HLN-fitted phase coherence lengths $L_\phi$ vs BTS thickness $d$.}
			\label{fig-Rvd}
			\end{figure}

The symmetric longitudinal magnetoresistance (MR) $R_{XX}$ in perpendicular magnetic field is shown in Fig-\ref{fig-Rvd}(a) for several BTS devices of different thicknesses. The MR shows a sharp cusp in low-field range, which is representative of weak antilocalization (WAL). It is a result of the negative interference in electron paths due to $\pi$ Berry's phase in TIs\cite{FuKane07}. The WAL effect is especially an indicator of topologically protected surface states as TIs belong to the symplectic class and unlike topologically trivial 2D systems, don't show a crossover to weak localization from WAL\cite{Chen_WAL_10}. Incidentally, fluctuations in the MR are also evident for thinner devices, which will be addressed later. The WAL correction to the conductivity has been computed by Hikami, Larkin and Nagaoka (HLN), and for the symplectic case it is\cite{HLN80}:
			\begin{equation}
			\begin{split}
			\delta G(B) &= G(B) - G(0)\\
			&\approx \alpha_\phi\frac{e^2}{2\pi\hbar}\left[ln\left(\frac{B_\phi}{B}\right)-\psi\left(\frac{1}{2}+\frac{B_\phi}{B}\right)\right]
			\end{split}
			\label{eq-redHLN}
			\end{equation}
Prefactor $\alpha_\phi$ in Eq-\ref{eq-redHLN} is indicative of the nature and number of conduction channels and $B_{\phi} \left(= \hbar/4eL_\phi^2\right)$ is the dephasing field, associated with the characteristic phase decoherence length $L_\phi$. Eq-\ref{eq-redHLN} is a simplified or reduced version of the full HLN conductivity correction, assuming strong spin orbit coupling in the transport direction, no magnetic scattering and large elastic scattering time\cite{HLN80}. $\alpha_\phi$ is exactly equal to $1/2\pi$ for the symplectic case of the 2D topological surface channel. Several of the device MR data are fitted to Eq-\ref{eq-redHLN} in low-field limit to extract $\alpha_\phi$ and $L_{\phi}$, as shown in Fig-\ref{fig-Rvd}(b). As can be seen from Fig-\ref{fig-Rvd}(c) for thinner devices $\alpha_\phi\in[1/2\pi,1/\pi]$ and for thicker ones it is larger than $1/\pi$. The exact meaning of the values and trends of the prefactor $\alpha_\phi$ has been a matter of some debate, and because of its empirical fitting nature it is more an indirect indicator of the underlying complex picture of multi-channel transport\cite{Chen_WAL_10,Steinberg_WAL_11,Chen_aWAL_11,Lee_BSTS_12,Xia_BSTS_13,Kim_Coherent_13}. Qualitatively however, one can distinguish regimes of transport: $\alpha_\phi$ can almost continuously vary from an ideal picture of parallel symplectic channels, i.e., surface states ($\nu/2\pi,\ \nu\in\mathbb{N}$), to a more complicated picture of surface states coupled via conductive bulk (non-integer multiples of $1/2\pi$). Value of $\alpha_\phi\in[1/2\pi,1/\pi]$ has been attributed to phase-preserving coherent scattering between the two surface states and bulk states, which are only partially decoupled such that the contribution effectively adds up to less than two full channels\cite{Steinberg_WAL_11,Chen_aWAL_11,Li_BTSUCF_12,Kim_Coherent_13,Yang_BSTGating_14}. Similarly, $\alpha_\phi>1/\pi$ may indicate a larger degree of separation of surfaces but with an addition of other channels: larger bulk contribution in thicker devices and trivial 2D subbands, occurring mainly due to surface band bending\cite{Bianchi_2DEG_10,Lee_BSTS_12,Yang_BSTGating_14}. This observation corroborates the $G_{\square}$ vs $d$ data from Fig-\ref{fig-RvT}(d). The argument is further supported by the phase coherence length data, as shown in Fig-\ref{fig-Rvd}(d) where $L_\phi$ is larger for thinner devices. This may be explained as a lower (higher) bulk channel contribution and hence a lower (higher) surface-to-bulk scattering in thinner (thicker) devices leading to a longer (shorter) phase-coherence time and length\cite{Steinberg_WAL_11}.
\vspace{-1em}
\subsection{Temperature-dependent Magnetotransport Measurements}
\label{sec-MRvT}
\begin{figure}[h]
				\includegraphics[width=0.5\textwidth]{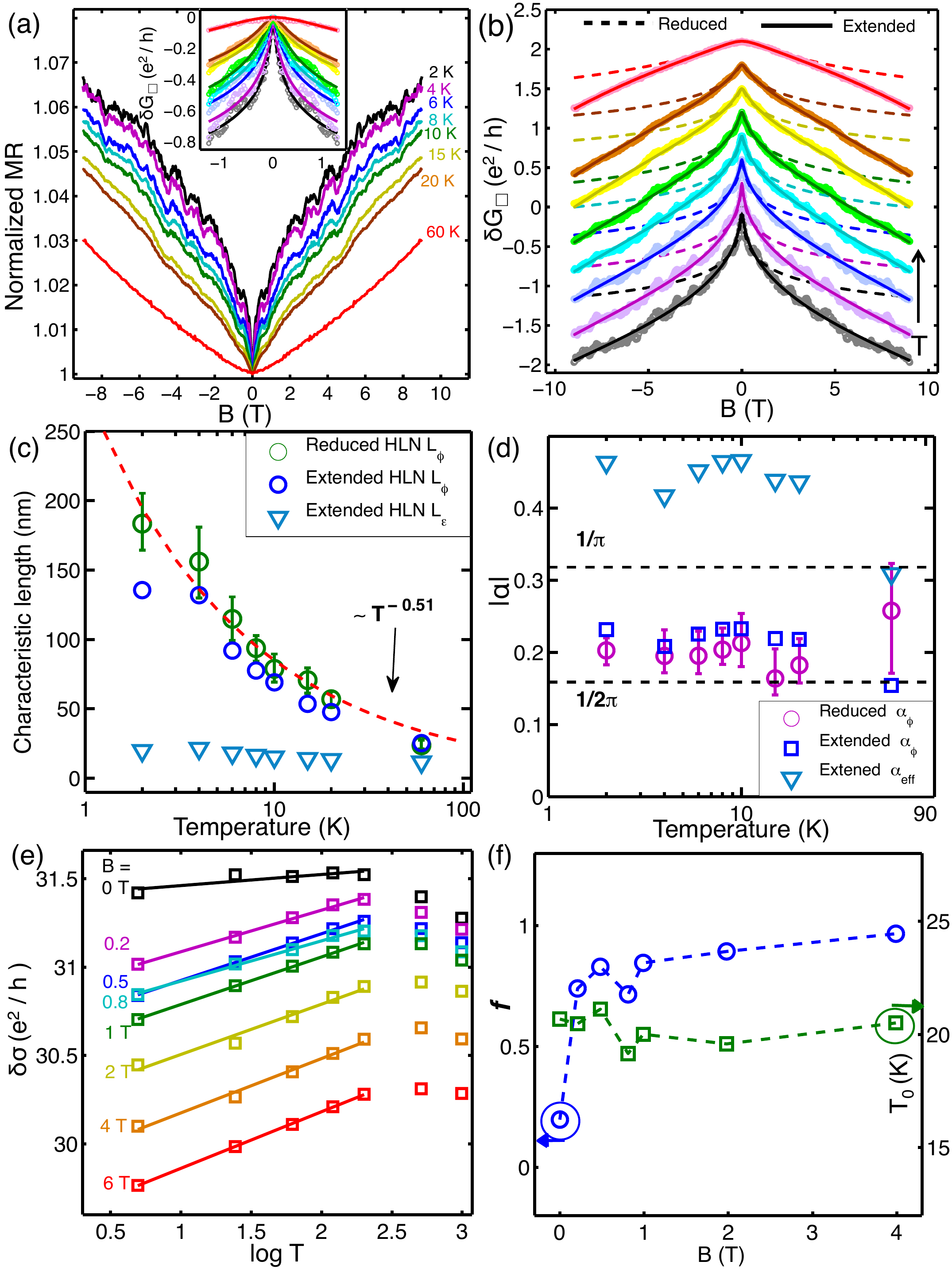}
				\caption{(a) Normalized symmetric longitudinal MR ($R_{XX}$) at different temperatures, in perpendicular magnetic fields. Inset shows reduced HLN fit to $\delta G_\square$ at low-fields. (b) Comparison of reduced and extended HLN fit to the full range $\delta G_\square$ data (curves shifted for clarity). (c) Characteristic lengths from the reduced and extended HLN fits, at different temperatures. Dashed line is a $\sim T^{-0.51}$ fit. (d) The prefactor $\alpha_\phi$'s from the reduced and extended HLN fits, at different temperatures (Extended HLN $\alpha_\phi = \alpha_{eff}/2$). (e) Linear dependence of the conductivity on logarithmic temperature in different magnetic fields, showing saturation behavior at high temperatures. (f) Coefficients $f$ and $T_0$ from the quantum correction the conductivity, fitted to Eq-\ref{eq-qcondgen}.}
			\label{fig-MRvT}
			\end{figure}

Temperature-dependent magnetoresistance measurement results for a candidate thin BTS device ($d=10$ nm) are shown in Fig-\ref{fig-MRvT}. The MR shows a sharp WAL cusp (Fig-\ref{fig-MRvT}(a)), which gets smaller as the sample temperature increases. The solid lines in Fig-\ref{fig-MRvT}(a) inset show the reduced HLN fits to the magnetoconductance at low magnetic fields. The limitation of the reduced fit is apparent if it is expanded to include full-range MR data (dashed lines in Fig-\ref{fig-MRvT}(b)), as the high-field magnetoconductivity is not dominated by the quantum-only correction of Eq-\ref{eq-redHLN} unlike in the low-field case.

\begin{figure}
				\includegraphics[width=0.5\textwidth]{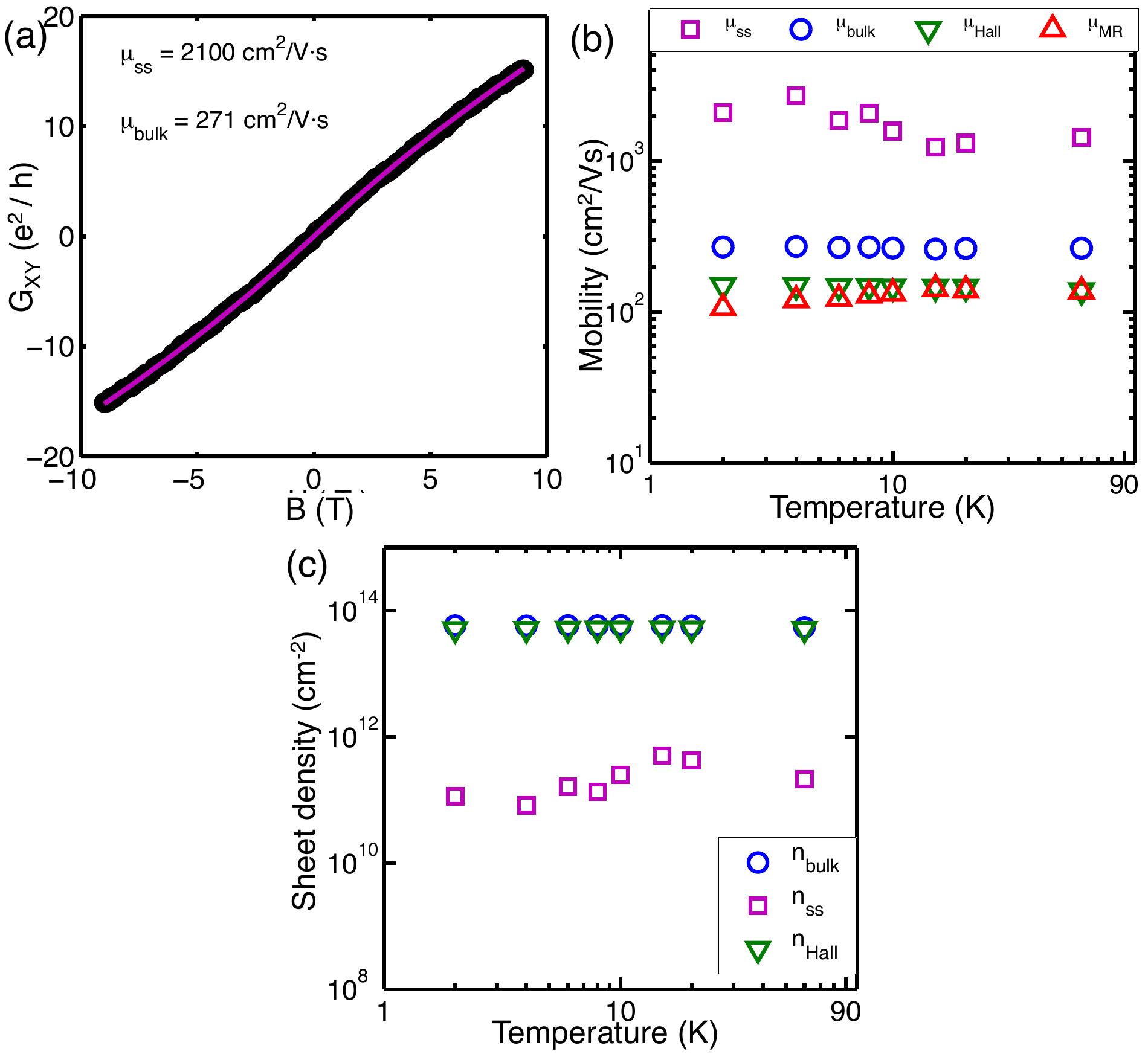}
				\caption{(a) Two-channel model fit applied to the Hall conductivity data at 2 K. (b) Mobilities from the two-channel model compared with the Hall and MR mobilities for different temperatures. (c) Sheet carrier densities from two-channel model compared with Hall concentration at different temperatures.}
				\label{fig-2chvT}
			\end{figure}

We have considered an extended version of the HLN equation as an alternative, with added terms:
			\begin{multline}
			\delta G(B) \approx \frac{\alpha_{eff}}{2}\frac{e^2}{2\pi\hbar}\left[ln\left(\frac{B_\phi}{B}\right) - \psi\left(\frac{1}{2}+\frac{B_\phi}{B}\right)\right] \\+ \alpha_{eff}\frac{e^2}{2\pi\hbar}\left[ln\left(\frac{B_\varepsilon}{B}\right) - \psi\left(\frac{1}{2}+\frac{B_\varepsilon}{B}\right)\right] - \beta B^2
			\label{eq-extHLN}
			\end{multline}
Eq-\ref{eq-extHLN} has two extra correction terms compared to Eq-\ref{eq-redHLN}. The second term is similar to the first in form, but it represents the contribution from elastic scattering\cite{HLN80,Zhang_60T_12,Dey14}. The prefactor $\alpha_{eff}$ is an indicator of multiple channels effectively contributing to the correction. The final term is the conventional quadratic cyclotron term, which provides a negative correction to the overall conductivity\cite{Assaf_extHLN_13,Shekhar_BTS_14}. The extended HLN fit of Eq-\ref{eq-extHLN} proves more reliable for full-range fitting (solid lines in Fig-\ref{fig-MRvT}(b)). The characteristic lengths associated with the two dephasing fields in Eq-\ref{eq-extHLN}, $L_\phi,\ L_\varepsilon$ and the reduced HLN $L_\phi$, are shown as a function of temperature in Fig-\ref{fig-MRvT}(c) matching closely. The dashed line represents a $T^{-0.51}$ dependence, which is an attribute of a 2D system and corresponds to Nyquist electron-electron decoherence\cite{Altshuler_EEI_82,Lee_BSTS_12,Dey14}. The elastic scattering length $L_\varepsilon<L_\phi$ and changes relatively little over the temperature range. $\alpha_\phi$ v $T$ from both the fits are shown in Fig-\ref{fig-MRvT}(d) matching relatively well. The equivalent prefactor of the surface channel in the extended HLN fit is obtained from the effective prefactor as $\alpha_\phi = \alpha_{eff}/2$. As before, $\alpha_\phi$ is slightly larger than $1/2\pi$ at lower temperatures for both fits, indicating the presence of mainly a symplectic 2D channel, partially decoupled with the bulk states, contributing to the WAL. It is also instructive to consider the value of $\pi\alpha_{eff}$, which is $\sim1.5$. A value larger than unity indicates the presence of more than one channel contributing\cite{Lee_BSTS_12}, especially with the elastic scattering term in Eq-\ref{eq-extHLN}. The data at 60K is almost parabolic with a very small WAL feature, such that the first and second terms in Eq-\ref{eq-extHLN} act equivalently for the purposes of the fit and gives larger error in the reduced HLN case. The fitting parameter $\beta$ can be expressed as $\mu_{MR}^2 G_\square(0)$, where $\mu_{MR}$ is the mobility estimated from the parabolic MR term and for isotropic scattering $\mu_{MR}$ should be approximately similar to $\mu_{Hall}$\cite{SchroderBook}. The $\mu_{MR}$ from the fit is very close to the Hall mobility (Fig-\ref{fig-2chvT}(b)), i.e., $\sim 150$ cm$^2$/V$\cdot$s. The temperature-dependent 2D behavior of $L_\phi$ and the values of $\alpha_\phi$ can be understood as arising from topological surface channel contributing to the WAL cusp, while the conductive bulk also contributes to the high-field MR behavior. In prior study on Bi$_2$Te$_3$ films, additional conductivity correction terms in the HLN equation from spin-orbit scattering, were also considered\cite{Zhang_60T_12,Dey14}. However, for our data the spin-orbit dephasing fields $B_{SOx,z}\ggg B_\phi,B_\varepsilon$, hence could be safely ignored from the fit. The contribution from the WAL effect dominates the conductivity at low-temperatures and at zero-field. The expected quantum correction to the temperature-dependent conductivity in disordered systems due to localization is (in units of $e^2/h$):
				\begin{equation}
					\delta\sigma = - \frac{\alpha_{qc}}{\pi}\ ln\left(\frac{\tau_\phi}{\tau}\right) = \frac{\alpha_{qc}p}{\pi}\ ln\left(\frac{T}{T_L}\right)
				\label{eq-qcond}
				\end{equation}
Where the phase coherence time $\tau_\phi\sim T^{-p}$ ($p=1$ for 2D), $\alpha_{qc}$ is a prefactor similar in nature to $\alpha_\phi$ from the HLN fit and $T_L$ is the temperature at which the correction disappears\cite{Beutler_EEI_88,Takagaki_CuBS_12,Lu_EEIWAL_14}. The conductivity should continue to increase with decreasing temperature for a purely WAL-like contribution. However, as discussed in Section \ref{sec-RvT}, the devices show a decrease in the conductivity with decreasing temperature, which is linearly proportional to logarithmic temperature and is attributed to EEI. The EEI correction is also logarithmic in nature similar to Eq-\ref{eq-qcond}, as seen from Eq-\ref{eq-EEI}. The low-temperature conductivity in different perpendicular magnetic fields can be then fitted to a generic equation of the form (in units of $e^2/h$):
				\begin{equation}
 						\delta\sigma = \frac{f}{\pi}\ ln\left(\frac{T}{T_0}\right)
				\label{eq-qcondgen}
				\end{equation}
Where $f$ is the slope of the line (equivalent to $\alpha_{qc}p$, $(1-\frac{3}{4}\tilde{F_\sigma})$ from Eq-\ref{eq-qcond} and Eq-\ref{eq-EEI}) and $T_0$ is the characteristic temperature. Fig-\ref{fig-MRvT}(e) shows the temperature dependent conductivity fits to Eq-\ref{eq-qcondgen}, and Fig-\ref{fig-MRvT}(f) shows the values of $f,\ T_0$ as a function of magnetic field. $f$ saturates to $\sim 1$ at fields higher than $2$ T, whereas $T_0$ is within $20\pm1$ K and approximately independent of field. Thus the WAL and EEI effects arise in a similar temperature range. The value of $\alpha_{qc}$ can be extracted from $\alpha_{qc} p=\delta f \approx 0.7$ with $p=1$ for 2D states, which is slightly different than the HLN-fitted $\pi\alpha_\phi \approx 0.66$ but still indicates effectively a single surface state with a bulk contribution\cite{Liu_EEI_11,Lu_EEIWAL_14}. This variation in $\alpha$ has been observed in previous experiments\cite{Beutler_EEI_88,Wang_EEI_11,Takagaki_CuBS_12,Roy_BT_13,Lu_EEIWAL_14}. The saturating value of $f$ at higher fields points to a dominant contribution from EEI, as the magnetic field dependence of EEI is weaker, while the low-field variation in $f$ may be attributed to the WAL effect, such as both effects act in combination for the BTS devices\cite{Beutler_EEI_88,Takagaki_CuBS_12,Takagaki_ST_12,Roy_BT_13,Lu_EEIWAL_14}.

Fig-\ref{fig-2chvT} shows the results obtained from a two-channel model\cite{Ren_BST_10,Steinberg_SS_10,Bansal_TI_12,Yang_BSTGating_14} fit to the Hall conductivity $G_{xy}$ data, to investigate the parallel contribution from the surface and bulk effective channels (see Appendix \ref{appsec-2ch} for details). Fig-\ref{fig-2chvT}(a) shows the two-channel fit to the Hall conductivity data for the 10 nm BTS device. The two mobilities and carrier densities extracted from the model fit are shown in Fig-\ref{fig-2chvT}(b) and (c), respectively, compared with the values computed directly from the Hall coefficient. The bulk carrier concentrations from both Hall and two-channel models are close to previously reported carrier densities for bulk crystals showing n-type doping\cite{Soonpaa_BTS_62,Ji_BTS_12}. The two-channel model reveals the presence of a higher mobility and lower carrier density surface channel, i.e., $\mu_{ss}, n_{ss}$, whereas the lower mobility and higher carrier density channel, $\mu_{bulk}, n_{bulk}$, is indicative of an effective contribution from the bulk channel. The two-channel model fits the data well up to higher temperatures, where the surface state channel is still found to contribute to the overall conduction, with the bulk conduction states always present.
 \vspace{-1em}
\subsection{Universal Conductance Fluctuations}
\label{sec-ucf}
			\begin{figure}
			\centering
				\includegraphics[width=0.5\textwidth]{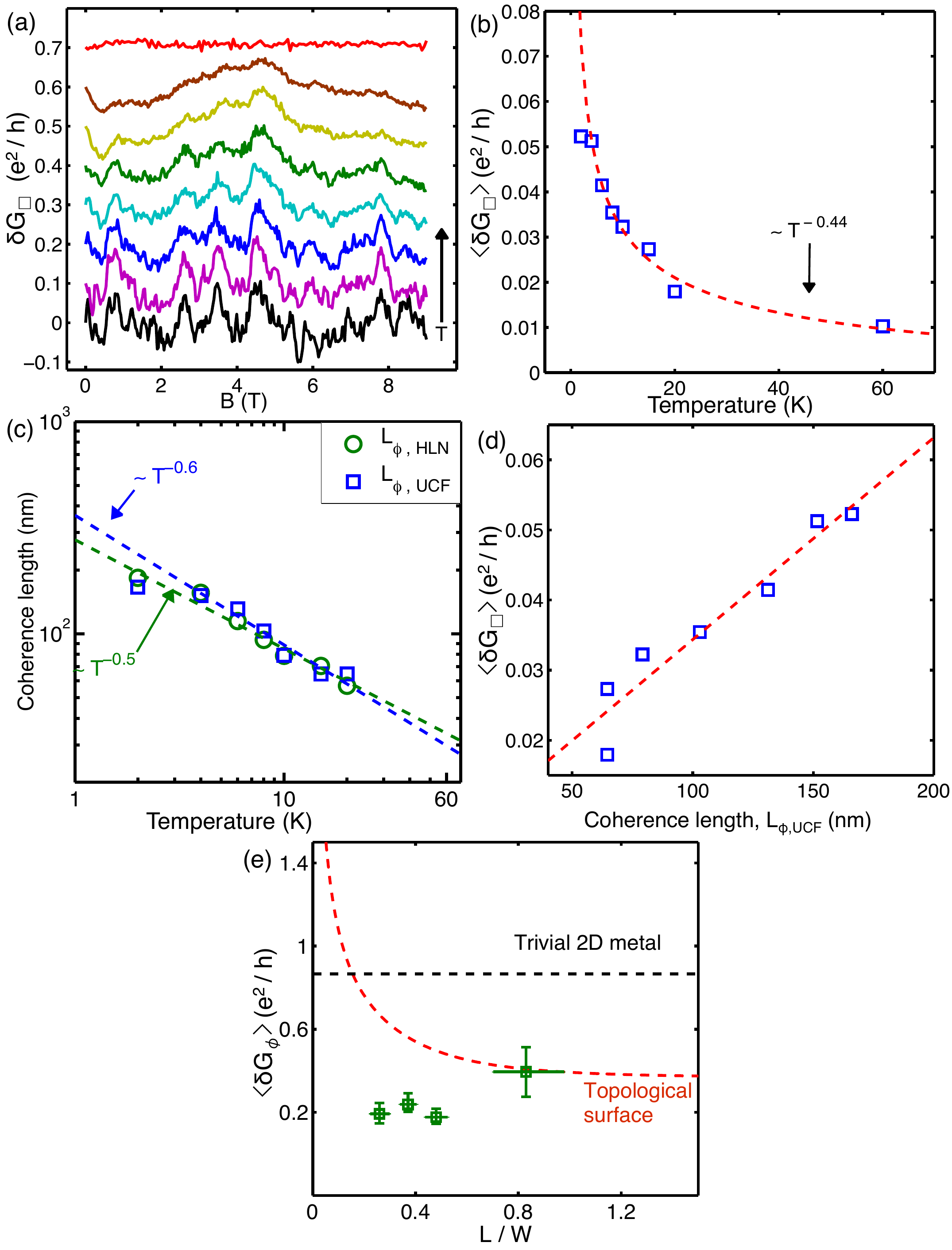}
				\caption{(a) Sheet magnetoconductance at different temperatures with smooth background subtracted. Curves are shifted for clarity. (b) Magnitude of the fluctuations as a function of temperature, showing a $\sim T^{-0.44}$ behavior expected from 2D UCF theory. (c) Phase coherence length $L_\phi$, as extracted from the UCF calculations, compared with the $L_\phi$ from HLN fitting. (d) RMS values of the fluctuations show an almost linear dependence on the corresponding phase coherence lengths, according to 2D UCF theory. (e) UCF magnitude in a phase-coherent box ($\langle\delta G_{\phi}\rangle$) in several devices as a function of their dimensional ratio $L/W$.}
			\label{fig-UCF}
			\end{figure}

Fluctuations in the magnetoresistance are evident for thinner devices, for example in the 10 nm device in Fig-\ref{fig-MRvT}(a) and can be visualized better by subtracting the smooth background from the extended HLN fits (Fig-\ref{fig-UCF}(a)), seen to be persisting up to higher temperatures. Universal conductance fluctuations (UCF) are a manifestation of an electron's path interfering with itself, as it goes around a defect site. If the phase of the electron is preserved over a mesoscopic scale $\mathcal{L}_{\phi}$, then the interference leads to measurable variance or fluctuations in the conductance. When the sample dimensions are larger than $\mathcal{L}_{\phi}$ there is some self-averaging due to changing impurity potential configuration\cite{Altshuler85,Lee_UCF_87}, which can reduce the overall amplitude of the fluctuations. To calculate the amplitude, a correlation function of the fluctuations is calculated as\cite{Lee_UCF_87}: $\mathcal{F}(\Delta B) = \langle\delta g(B)\cdot\delta g(B+\Delta B)\rangle$, where $\delta g = g(B) - \langle g(B)\rangle$ (see Appendix \ref{appsec-ucf} for examples of the correlation function). The UCF magnitude is obtained at $\sqrt{\mathcal{F}(0)}$ and is of the order of $\sim 0.05\ e^2/h$, decaying as $T^{-0.44}$ from our experimental data (Fig-\ref{fig-UCF}(c)). This temperature dependence is similar to theoretically expected $1/\sqrt{T}$ dependence in a 2D system\cite{Altshuler85,Lee_UCF_87,Lee_BSTS_12,Li_BTSUCF_12,Xia_BSTS_13}. The correlation-field $B_c$ (related to the scale of the UCF) can be calculated from the correlation function as $\mathcal{F}(B_c) = \frac{1}{2}\mathcal{F}(0)$ and the related phase coherence length as $L_{\phi,\ UCF}=\sqrt{\Phi_0/B_c}$, where $\Phi_0$ is the flux quantum\cite{Lee_UCF_87}. As seen in Fig-\ref{fig-UCF}(c) $L_{\phi,\ UCF}$ matches closely with $L_{\phi,\ HLN}$ and decays as $\sim T^{-0.6}$, which confirms that the fluctuations are primarily from the 2D surface channel\cite{Lee_UCF_87}. Also expected from 2D UCF theory is the linear relation of rms value of the fluctuations with the corresponding phase coherence length\cite{Lee_UCF_87,Gehring_Kawazulite_13} (see Fig-\ref{fig-UCF}(d)). To see the consequence of finite size effects on the UCF magnitudes, devices of different dimensions ($L,\ W$) can be compared for their rms values. For a true comparison the rms value of the UCF within a phase-coherent box $\mathcal{L}_\phi\times\mathcal{L}_\phi$ should be considered, which calculated as $\langle\delta G_\phi\rangle = \sqrt{N}\frac{L}{W}\langle\delta G_{\square}\rangle$, where $N=L W/L_\phi^2$ is the number of phase coherent boxes in a $L\times W$ sample\cite{Pal_UCF_12}. Rossi \textit{et al.} have proposed an approach to compare UCF amplitudes in Dirac materials, independent of impurity density, disorder strength and correlation length\cite{Rossi_UCF_12}:
			
			\begin{equation}
				\langle\delta G_\phi^2\rangle = \left(\frac{e^2}{\pi^2h}\right)^2 \sum_{n_x=1,n_y=-\infty}^{\infty} \frac{12g_s g_v}{\left(n_x^2+4\left(\frac{L}{W}\right)^2n_y^2\right)^2}
			\label{eq-UCF}
			\end{equation}

In Eq-\ref{eq-UCF} the spin and valley degeneracies for TI surface channel is $g_s g_v=1$\cite{Rossi_UCF_12}. Fig-\ref{fig-UCF}(e) shows the phase coherent UCF magnitudes of several devices as a function of the dimensional ratio $L/W$. The red dotted line shows the behavior of a topological surface channel from Eq-\ref{eq-UCF}; the black dotted line is the expected value of the UCF magnitude ($\approx\ 0.86\ e^2/h$) for a topologically trivial 2D-electron gas (2DEG)\cite{Altshuler_exUCF_85,Li_BTSUCF_14}. The UCF magnitudes are somewhat less than the magnitude expected from topological states according to Eq-\ref{eq-UCF}, but they are much less than trivial 2D metal values, indicating that the conductivity fluctuations largely arise due to the topological surface states\cite{Li_BTSUCF_14,Gehring_Aleksite_15}. 
 \vspace{-1em}
\section{Conclusions}
\label{sec-summary}
In summary, we demonstrate van der Waals epitaxial growth of crystalline Bismuth Telluro-Sulfide (BTS) nanosheets on SiO$_2$ and mica substrates. As grown BTS material is obtained in highly layered, good-quality crystalline nanosheets. Detailed transport experiments in devices of BTS indicate the presence of surface states, albeit with bulk states still present in the transport. Weak anti-localization and universal conductance fluctuation signatures are seen in the magnetoresistance of the BTS devices. Evidence of a combination of both weak antilocalization and electron-electron interaction effects is seen from analyzing the insulating ground state in the temperature and field-dependent conductivity data. An extended-HLN model is considered, which provides excellent fitting to longitudinal magnetoconductance data in high-field range, and indicates the presence of 2D surface states, partially coupled to the bulk conducting states. Two-channel Hall conductivity model in conjunction with the extended-HLN model provides a good fit for the magnetoconductance data. The HLN effective prefactor $\alpha_{eff}$ provides a good qualitative tool to understand the regimes of transport in the device, and its value is indicative of the contribution of different channels acting in parallel. Universal conductance fluctuations are observed in thin BTS devices. The temperature dependent phase coherence lengths from UCF data are in reasonable agreement with those from WAL data, showing 2D behavior arising from topological surface states. The magnetoconductivity data and its empirical modeling show that the bulk conduction channels are still present in the BTS material, likely due to high level of n-type doping by chalcogen deficiency. Our studies confirm BTS as a candidate 3D TI material in conjunction with the previous APRES measurements, and takes a step closer to understanding transport mechanism in TI materials. Future growth experiments of the BTS material on highly crystalline substrates and substitutional doping with more p-type elements such as Sb, are expected to improve the electronic properties of the BTS material system by reducing the chalcogen deficiency doping and pushing the Fermi level further into the bandgap, leading to the predicted promising nature of the Sulfur based ternary tetradymite. Additionally, more experiments with top and bottom gating on high-quality BTS material are also expected to reveal further the nature and contribution of the multiple transport channels acting in parallel in 3D TIs. BTS provides an alternative material with surface-states accessible by transport measurements, to further probe the topological nature of the states and for potential heterostructure-based applications of different 3D TIs in nanoelectronics. 
 \vspace{-1em}
\begin{acknowledgements}
This work was supported in part by the Nanoelectronics Research Initiative's (NRI) South West Academy of Nanoelectronics (SWAN) and the National Nanotechnology Coordinated Infrastructure (NNCI). T.T. thanks Kyounghwan Kim for helping to build and troubleshoot the growth system, Dr. Anupam Roy and Rik Dey for valuable discussions and Dr. Babak Fallahazad for helping with wirebonding.
\end{acknowledgements}
\vspace{-1em}
\appendix
\section{Electron-electron Interaction Driven Insulating Ground State}
\label{appsec-EEI}
 \vspace{-1em}
The sheet resistance of the BTS devices shows indications of an insulating ground state, as observed in the main text. Oftentimes, the insulating ground state manifests as an increase (decrease) in the sheet resistance (conductance) as the sample is further cooled down. The correction to the conductivity due to the electron-electron interaction (EEI) effects in 2D systems is given by Eq-\ref{eq-EEI} in the main text. The temperature dependent conductance data of several BTS devices fit the linear expression of Eq-\ref{eq-EEI} in $ln\ T$, as seen from Fig-\ref{fig-RvTEEI}. Assuming that the total conductivity can be represented as a sum of effectively two types of contribution, i.e., surface and bulk channels\cite{Steinberg_SS_10,Ren_BST_10}:

	\begin{equation}\label{appeq-EEI2ch}
			\sigma = \sigma_b + \frac{G^{ss}_\square}{d}
	\end{equation}

Where $\sigma, \sigma_b$ and $G^{ss}_\square$ are the total conductivity, bulk conductivity and surface state (SS) conductance, respectively. Assuming that the bulk conductivity component is largely independent of the thickness of the film, and that the SS conductance follows the 2D EEI relation of Eq-\ref{appeq-EEI2ch} (up to some correction factor converting between conductance and conductivity), a temperature-dependent conductivity correction can be written as:
	\begin{equation}\label{appeq-EEITmin}
		\begin{split}
			\frac{\partial\sigma}{\partial T} &\approx \frac{\partial\sigma_b}{\partial T} + \frac{C}{d\cdot T}\\
			\mbox{At the resistance minima, }&T=T_{min},\ \frac{\partial\sigma}{\partial T} = 0\\
			\therefore 0 &\approx \frac{\partial\sigma_b}{\partial T}\bigg|_{T_{min}} + \frac{C}{d\cdot T_{min}}\\
			\therefore T_{min} &\approx \bigg(\frac{-C}{\frac{\partial\sigma_b}{\partial T}\big|_{T_{min}}}\bigg)\cdot\frac{1}{d}
		\end{split}
	\end{equation} 

$C$ in Eq-\ref{appeq-EEITmin} is a combined constant factor of all the temperature independent variables obtained after differentiating the expression in Eq-\ref{appeq-EEI2ch} and Eq-\ref{eq-EEI}. As can be seen from Eq-\ref{appeq-EEITmin}, the temperature of the resistance minima (or conductivity maxima) scale roughly as $1/d$. This thickness dependent behavior is noted in the main text for BTS devices, where a $\sim d^{-0.94}$ fit is obtained for the experimentally observed $T_{min}$ for several devices.
\vspace{-1em}
\section{Two-channel Model}
\vspace{-1em}
\label{appsec-2ch}

A two-channel model is widely used to fit Hall conductance data in TI devices, as also utilized in Appendix \ref{appsec-EEI}\cite{Ren_BST_10,Steinberg_SS_10,Bansal_TI_12}. Fig-\ref{fig-2chvT} in main text shows an example two-channel fit to a candidate device ($d=10$ nm). A generic multi-carrier model can be represented as:
	\begin{equation}
		G_{XX} = e\sum_i\frac{n_i\mu_i}{1+\mu_i^2B^2},\ G_{XY} = eB\sum_i\frac{n_i\mu_i^2}{1+\mu_i^2B^2}
	\label{appeq-2ch-gen}
	\end{equation}
Where $n_i,\ \mu_i$ are the carrier concentration and mobility of the $i^{\mbox{th}}$ channel, respectively. $G_{XX,\ XY}$ are the conductance tensor components. With some basic algebra and limiting assumptions the number of unknown variables in the fit can be reduced and the following equation can be used for a two-channel model\cite{Bansal_TI_12}:
	\begin{equation*}\label{appeq-2ch-fit}
		\begin{split}
		&G_{XY}\\
		= &eB\left(\frac{k_1\mu_1-k_2}{\left(\frac{\mu_1}{\mu_2} - 1\right)\cdot(1+\mu_2^2B^2)}+\frac{k_1\mu_2-k_2}{\left(\frac{\mu_2}{\mu_1} - 1\right)\cdot(1+\mu_1^2B^2)}\right)\\
		k_1 & = G_{XX}(0)/e,\ k_2 = \lim_{B\to0} G_{XY}(B)/eB\\
		n_1 & = \frac{k_1\mu_2-k_2}{\mu_1\mu_2-\mu_1^2},\ n_2 = \frac{k_1\mu_1-k_2}{\mu_1\mu_2-\mu_2^2}
		\end{split}
	\end{equation*}
\vspace{-1em}
\section{Universal Conductance Fluctuations Correlation Function}
\label{appsec-ucf}
\begin{figure}[h]
		\centering
		\includegraphics[width=0.45\textwidth]{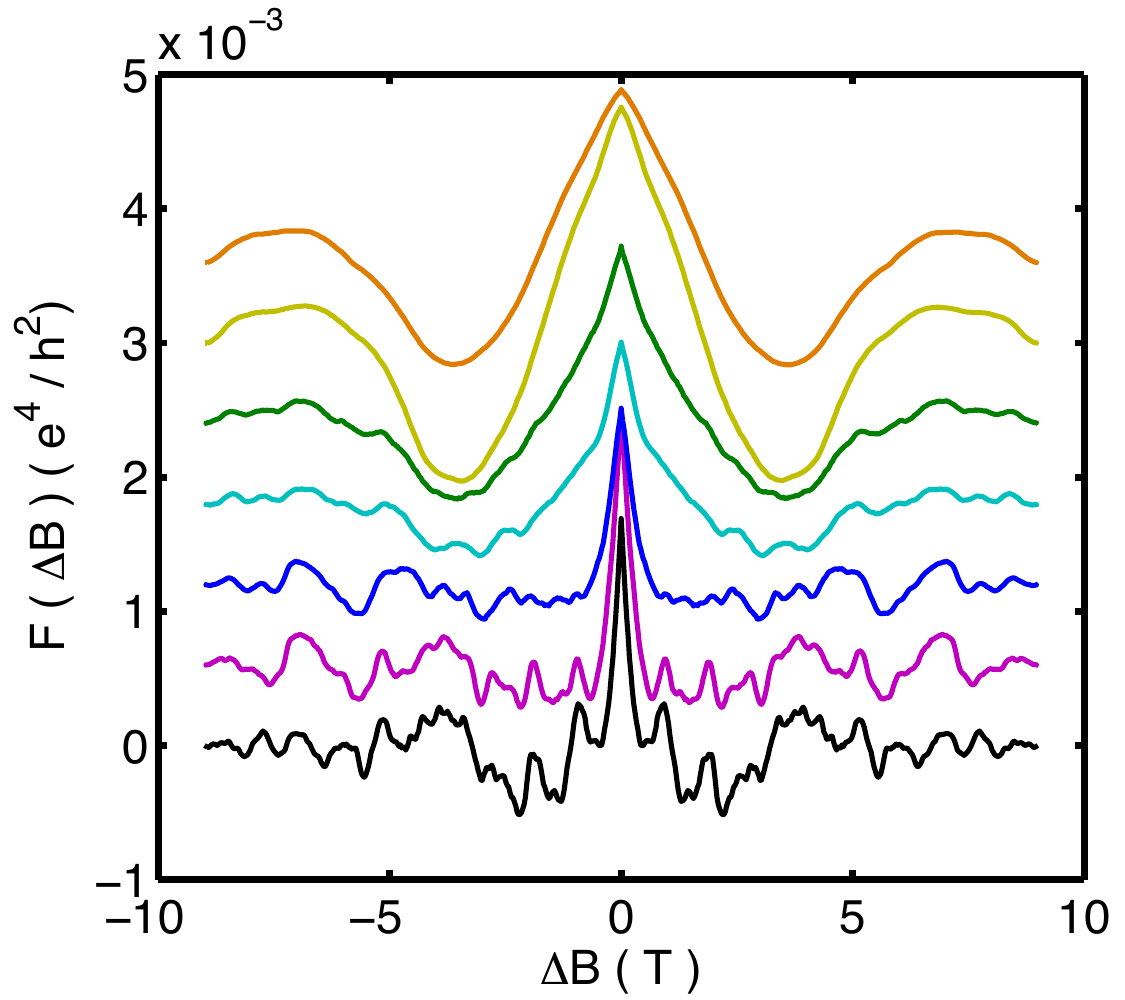}
		\caption{Correlation functions for universal conductance fluctuations in the 10 nm device, at different temperatures. The color scheme is consistent with the main text. Curves are shifted for clarity.}
		\label{appfig-ucf}
	\end{figure}

\vspace{-1em}
The magnitude of the conductance fluctuations and the correlation field $B_c$ can be calculated from the correlation function, in a device exhibiting UCF in the magnetoresistance. The correlation function can be calculated as: $\mathcal{F}(\Delta B) = \langle\delta g(B)\cdot\delta g(B+\Delta B)\rangle$, where $\delta g = g(B) - \langle g(B)\rangle$. Examples of the correlation function for a candidate device (10 nm) are shown in Fig-\ref{appfig-ucf}.
\bibliography{tkt_BTS_TI-refs}

\end{document}